\documentclass{ifacconf}

\usepackage{graphicx}      % include this line if your document contains figures
\usepackage{natbib}        % required for bibliography
\usepackage[overload]{empheq}
\usepackage{amsmath,amsfonts}
\usepackage{mathtools}
\usepackage{bm}
\usepackage{xcolor}
\usepackage{tikz}
\usepackage{subcaption}
\usepackage{stmaryrd}
\usepackage{bm}
\newcommand{\kr}{\odot}
\newcommand*\rfrac[2]{{}^{#1}\!/_{#2}}
\newenvironment{review}{\color{black}}{}
\newcommand{\rev}[1]{\begin{review}#1\end{review}}

\newenvironment{revieww}{\color{black}}{}
\newcommand{\revv}[1]{\begin{revieww}#1\end{revieww}}
\begin{document}
\begin{frontmatter}

\title{Decoupling P-NARX models using filtered CPD} 

\thanks[footnoteinfo]{This work was supported by the Flemish fund for scientific research FWO under license number G0068.18N.}

\author[First]{Jan Decuyper} 
\author[Second]{David Westwick} 
\author[Third]{Kiana Karami}
\author[First,Fourth]{Johan Schoukens}

\address[First]{Department of Engineering Technology, Vrije Universiteit Brussel, Pleinlaan 2, 1050 Brussels, Belgium (e-mail: jan.decuyper@vub.be, johan.schoukens@vub.be)}
\address[Second]{Department of Electrical and Computer Engineering, University of Calgary, Calgary, Canada (e-mail: dwestwic@ucalgary.ca)}
\address[Third]{Department of Electrical Engineering and Electrical Engineering Technology, Penn state Harrisburg, Middletown, PA, USA, 17057 (e-mail: kfk5551@psu.edu)}
\address[Fourth]{Department of Electrical Engineering, Eindhoven University of Technology, Eindhoven, The Netherlands}
 %  Seoul National University, Seoul, Korea, (e-mail: author@snu.ac.kr)}

\begin{abstract}
\revv{Nonlinear Auto-Regressive eXogenous input (NARX) models are a popular class of nonlinear dynamical models. Often a polynomial basis expansion is used to describe the internal multivariate nonlinear mapping (P-NARX). Resorting to fixed basis functions is convenient since it results in a closed form solution of the estimation problem. The drawback, however, is that the predefined basis does not necessarily lead to a sparse representation of the relationship, typically resulting in very large numbers of parameters. So-called decoupling techniques were specifically designed to reduce large multivariate functions. It was found that, often, a more efficient parameterisation can be retrieved by rotating towards a new basis. Characteristic to the decoupled structure is that, expressed in the new basis, the relationship is structured such that only single-input single-output nonlinear functions are required. Classical decoupling techniques are unfit to deal with the case of single-output NARX models. In this work, this limitation is overcome by adopting the filtered CPD decoupling method of \cite{decuyperFCPD2021}. The approach is illustrated on data from the Sliverbox benchmark: measurement data from an electronic circuit implementation of a forced Duffing oscillator.}    
   % Abstract of not more than 250 words.
%The polynomial decoupling algorithm was originally developed for Multiple-Input Multiple-Output (MIMO) polynomials, and involved computing the Canonical Polyadic Decomposition (CPD) of the Jacobian tensor. The method relied intrinsically on the uniqueness properties of the tensor decomposition. One of the resulting factor matrices, which contains information about the derivatives of the decoupled polynomials, resulted in extremely noisy estimates of these functions, whenever non-unique decompositions were encountered. One possibility is to impose a polynomial constraint on the CPD, but this results in a non-convex optimisation.
%\cite{decuyperFCPD2021} have proposed a non-parametric filtered CPD that uses regularisation to impose a smooth solution of this matrix factor.
%Computing the Jacobian of a Multiple-Input Single-Output function, such as a polynomial NARX model,  produces a matrix, rather than a tensor, thus the uniqueness properties of the CPD are lost. The non-parametric filtered CPD provides an alternative that allows one to avoid the non-convex optimisation resulting from imposing polynomial constraints on the CPD.  This approach is compared to the use of polynomial constraints in decoupling the nonlinearities in polynomial NARX models.  Both approaches are applied to data from the Sliverbox benchmark: measurement data from an electronic circuit implementation of a forced Duffing oscillator. 
\end{abstract}


\end{frontmatter}
%===============================================================================

\section{Introduction}

Nonlinear auto-regressive exogenous input (NARX) models have been extensively used to describe nonlinear systems. It is a black-box system identification technique which has proven to be useful in a wide range of applications \citep{chan2015,zhao2013}. NARX models describe dynamical nonlinear behaviour by relating the current output sample to both past output samples, and current and past input samples. Defining 
\begin{multline}
\label{e:x}
\boldsymbol{x} = \{u(t), u(t-1), \dots, u(t-n_u), \\ 
 y(t-1), \dots, y(t-n_y)\}, \quad \quad
\end{multline}
a general single-output NARX model is described by \citep{billings2013}
\begin{equation}
y(t) = F(\boldsymbol{x})+e(t),
\end{equation}
where $F: \mathbb{R}^{n_u + n_y +1} \rightarrow \mathbb{R}$ is a static multiple-input single-output (MISO) nonlinear function and $e(t)$ is an equation error which is assumed to be a sequence of independent identically distributed (IID) random variables. The function $F$ may be described by any class of functions, e.g.\ it can be a neural network or a wavelet network. Often, however, a basis expansion is preferred. In that case, a direct estimate of the model parameters follows from linear regression by using the measured outputs in the regressor (minimising the equation error). A popular choice is the polynomial basis, leading to so-called Polynomial-NARX or P-NARX models.

The ease of identification is however countered by a number of fundamental disadvantages. P-NARX models are notorious for the number of parameters that have to be identified, growing both with the number of past inputs and past outputs, $n = n_u + n_y + 1$, and combinatorially with the nonlinear degree $d$. Moreover, being a black-box method, the obtained model is \revv{often} very hard to interpret. \revv{General practice is to use regularisation to steer the optimisation towards meaningful terms and hence reduce the number of parameters. Often, however, no sparse representation can be obtained using the classical monomial basis. The present work promotes the use of decoupling techniques, which were designed to translate the relationship into a more favourable basis.}  
%, a total number of $\left(\begin{matrix} d+n \\ n \end{matrix} \right)$ is obtained

Similar issues have already been addressed in the context of nonlinear state-space models, where typically a multiple-input multiple-output (MIMO) polynomial is used to describe the nonlinearity. It was found that by \emph{decoupling} the multivariate polynomial into a number of univariate polynomials, both insight and model reduction could be obtained \citep{decuyper2021}. 

In general, decoupling techniques aim at transforming generic multivariate nonlinear functions into decoupled functions. The decoupled structure is characterised by the fact that the nonlinear relationship is described by a number of univariate functions of intermediate variables. Given a generic nonlinear function
\begin{equation}
\label{e:1}
\boldsymbol{q} = \boldsymbol{f}(\boldsymbol{p})
\end{equation}
with $\boldsymbol{q} \in \mathbb{R}^m$ and $\boldsymbol{p} \in \mathbb{R}^n$, the idea is to introduce an appropriate linear transformation of $\boldsymbol{p}$, denoted $\boldsymbol{V}$, such that in this alternative basis, univariate functions may be used to describe the nonlinear mapping. The decoupled function is then of the following form
\begin{equation}
\label{e:2}
\boldsymbol{f}(\boldsymbol{p}) = \boldsymbol{W} \boldsymbol{g}(\boldsymbol{V}^{\text{T}} \boldsymbol{p})
\end{equation}
where the $i$th function is $g_i(z_i)$ with $z_i = \boldsymbol{v}_i^{\text{T}} \boldsymbol{p}$, emphasising that all functions are strictly univariate. The number of allowed univariate functions, denoted $r$, is a user choice which can be used to control the model complexity. \revv{In some cases prior knowledge of the system may dictate that a certain number of nonlinear components drive the nonlinearity. Whenever no such knowledge is available, a scan over $r$ is performed.} The number of univariate functions plays a crucial role since it will determine whether the implied equivalence of Eq.~\eqref{e:2} exists. A second linear transformation $\boldsymbol{W}$, maps the function back onto the outputs. The linear transformations then have the following dimensions: $\boldsymbol{V} \in \mathbb{R}^{n \times r}$ and $\boldsymbol{W} \in \mathbb{R}^{m \times r}$.

The original decoupling procedure requires the function to be of the MIMO-type \citep{dreesen2014}, i.e\ $m>1$, excluding the class of single-output P-NARX models. The method is based on the canonical polyadic decomposition (CPD) of a 3-way tensor, constructed out of evaluations of the Jacobian matrix along a number of operating points (see Section \ref{s:FCPD}). It exploits the uniqueness properties of the CPD in order to retrieve estimates of the univariate functions $g_i$. In \cite{westwick2018}, the issue with single-output functions was circumvented by resorting to evaluations of the Hessian, again retrieving a 3-way tensor. In practice, however, the uniqueness of the CPD is often not guaranteed \citep{decuyper2019}, leading to very noisy estimates of $g_i$ or its derivatives. As a solution, \cite{karami2021} proposed to add polynomial constraints when factoring the Hessian. Accurate decoupled models could be obtained, although at the price of computing the Hessian.

In this work, the filtered CPD approach of \cite{decuyperFCPD2021} will be used. It acts on the basis of first order derivative information, avoiding the computation of the Hessian, and resorts to non-parametric soft constraints to ensure meaningful estimates of $g_i$ are obtained.

\section{ Filtered CPD decoupling}
\label{s:FCPD}
\begin{figure}
\begin{center}
\includegraphics[width=0.47\textwidth]{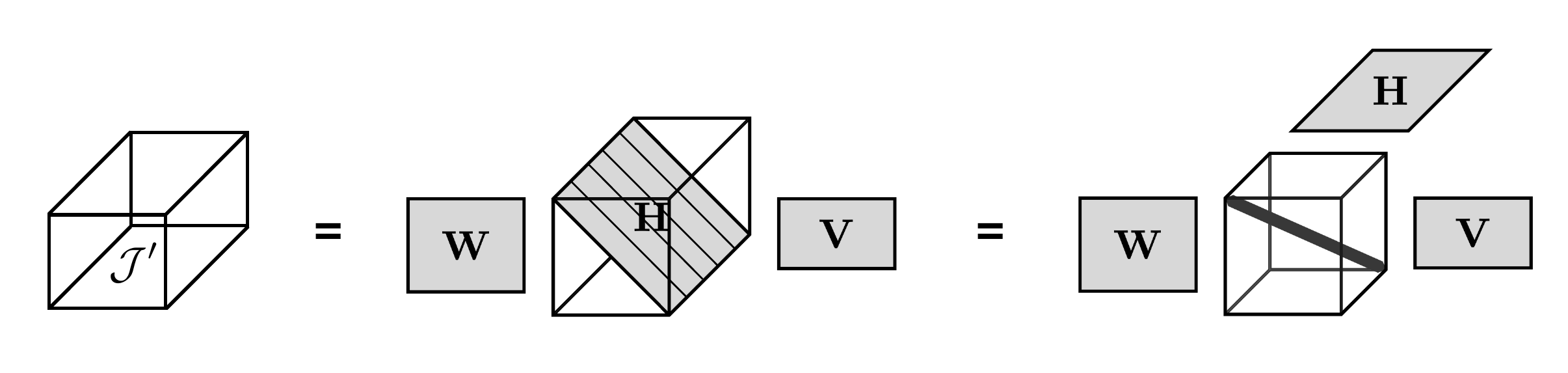}
\caption{Centre: a collection of evaluations of the Jacobian of the decoupled function (Eq.~\eqref{e:2}), stacked in the third dimension. Left: the corresponding third order tensor, Right: extracting the central diagonal plane reveals a diagonal tensor decomposition.}
\label{f:1}
\end{center}
\end{figure}

The filtered CPD approach (F-CPD) links the original function to a decoupled function on the basis of its first order derivative information. The method relies on the underlying diagonal structure of the Jacobian (Eq.~\eqref{e:3}), which follows from the use of univariate functions $g_i$. Denoting the left and right hand side Jacobian of Eq.~\eqref{e:2} by $\boldsymbol{J}$ and $\boldsymbol{J}'$, respectively, we have that
\begin{equation}
\label{e:3}
\boldsymbol{J}' = \boldsymbol{W} \operatorname{diag}\left(\begin{bmatrix} h_1(z_1) \quad \cdots \quad h_r(z_r) \end{bmatrix}\right) \boldsymbol{V}^{\text{T}}
\end{equation}
in which case $h_i(z_i) \coloneqq \frac{dg_i(z_i)}{dz_i}$ represents the derivative of the univariate function $g_i(z_i)$ with respect to its argument. Evaluating this Jacobian in $N$ operating points, i.e.\ for $\{\boldsymbol{p}[1], \cdots, \boldsymbol{p}[N]\}$, and collecting the evaluations in a third dimension, expands the object into a three-way array $\mathcal{J}' \in \mathbb{R}^{n \times m \times N}$. This is illustrated graphically in Fig.~\ref{f:1}. Notice that the diagonal plane, $\boldsymbol{H} \in \mathbb{R}^{N \times r}$, stores evaluations of $h_i$, i.e\ the derivative of the functions $g_i$. Given the diagonal form, the collection of Jacobians may be written as a sum of $r$ outer products (or rank-one terms). Element-wise we have that,
\revv{
\begin{equation}
%\begin{aligned}
\label{e:outer}
%\color{blue} \mathcal{J} = \sum_{i=1}^r \boldsymbol{w}_i \circ \boldsymbol{v}_i \circ \boldsymbol{h}_i,  \color{black}\\
%\mathcal{J}'_{jkl} = \sum_{i=1}^r \sum_{j=1}^m \sum_{k=1}^n \sum_{l=1}^N w_{ji}v_{ki}h_{li},
\mathcal{J}'_{\left[s,k,l\right]} = \sum_{i=1}^r w_{si}~v_{ki}~h_{li},
%\end{aligned}
\end{equation}
$\text{for}~s=1,\dots,m~\text{and}~k=1,\dots,n~\text{and}~l=1,\dots,N$.}  A sum of rank-one terms defines a diagonal tensor decomposition \citep{kolda2009}. 
%\rev{where lower bold faced letters indicate column vectors and $\circ$ denotes the outer product.} Extracting the central plane reveals a diagonal core tensor with diagonal elements equal to one. 
The latter is illustrated on the right in Fig.~\ref{f:1}. %It follows that $\mathcal{J}'$ is a tensor of size $\mathcal{J}' \in \mathbb{R}^{n \times m \times N}$.
%The decomposed tensor may be written in shorthand notation
%\begin{equation}
%\mathcal{J}' = \llbracket \boldsymbol{W}, \boldsymbol{V}, \boldsymbol{H} \rrbracket
%\end{equation}

\cite{dreesen2014} found that the underlying diagonal form of $\mathcal{J}'$ can be exploited in the decoupling process. It was suggested to construct a third order tensor, $\mathcal{J}$, out of evaluations of the Jacobian of the known function, $\boldsymbol{f}(\boldsymbol{p})$, and compute a diagonal decomposition such that $\mathcal{J} \approx \mathcal{J}'$. In the filtered CPD approach, finite difference filters are introduced into the decomposition. This allows for the Jocobian tensor to be decomposed into the more convenient factors $\{ \boldsymbol{W}, \boldsymbol{V}, \boldsymbol{G}\}$, where $\boldsymbol{G}$ stores evaluations of the univariate functions $g_i$. 

%Noticing that the right hand side of Fig.~\ref{f:1} corresponds to a diagonal tensor decomposition, and recollecting that in order for the equality of Eq~\eqref{e:2} to hold, the Jacobians need to match, forms the basis of the decoupling algorithm of \cite{dreesen2014}. 
The method of \cite{decuyperFCPD2021} can be summarised in three steps:
\begin{enumerate}
\item Collect the Jacobian matrices of the known function, $\boldsymbol{J}$, and stack them into a three-way array, i.e.\ the Jacobian tensor $\mathcal{J}  \in \mathbb{R}^{n \times m \times N}$.
\item Factor $\mathcal{J}$ into $\{\boldsymbol{W}, \boldsymbol{V}, \boldsymbol{G}\}$ by computing a filtered diagonal tensor decomposition (F-CPD). 
\item Retrieve the functions, $g_i$, by parametrising the nonparametric estimates stored in $\boldsymbol{G}$.
\end{enumerate}

At the core of the algorithm lies an alternating least-squares routine which iteratively updates $\boldsymbol{W}, \boldsymbol{V}$, and $\boldsymbol{G}$. \revv{Starting from a random initialisation\footnote{\revv{The retrieved local minimum depends on the initialisation point. In this work, however, no significant impact of the initialisation was observed when tackling the benchmark problem of Section \ref{s:benchmark}.}} the following norm is progressively minimised} 
\begin{equation}
\label{e:norm}
\underset{\boldsymbol{W},\boldsymbol{V},\boldsymbol{G}}{\operatorname{arg~min}}~\lVert \mathcal{J}  - \llbracket \boldsymbol{W}, \boldsymbol{V}, \mathcal{F}_C(\boldsymbol{V}) \circ \boldsymbol{G} \rrbracket \rVert_F^2,
\end{equation}
where the shorthand notation $\mathcal{J}' =  \llbracket \boldsymbol{W}, \boldsymbol{V}, \mathcal{F}_C(\boldsymbol{V}) \circ \boldsymbol{G} \rrbracket$ is used \rev{and `$\circ$' defines a matrix-column product, illustrated by Eq.~\eqref{e:hg}}. In this formulation, $\mathcal{F}_C \in \mathbb{R}^{N \times N \times r}$ stores a collection of \emph{finite difference filters}, with the $i$th filter
\begin{equation}
\boldsymbol{F}_{C_i} \coloneqq \mathcal{F}_{C_{[:,:,i]}}.
\end{equation}
A finite difference filter is a matrix which upon multiplication with a vector of function evaluations, returns a finite difference approximation. Recall that the $i$th column of $\boldsymbol{H}$ stores evaluations of the derivative of $g_i$ with respect to $z_i = \boldsymbol{v}_i^{\text{T}} \boldsymbol{p}$, resulting in the dependency $\mathcal{F}_C(\boldsymbol{V})$. We may then express $\boldsymbol{h}_i$, as the finite difference of the $i$th column of $\boldsymbol{G}$,
\begin{equation}
\label{e:hg}
\boldsymbol{h}_i \coloneqq \boldsymbol{F}_{C_i} \boldsymbol{g}_i.
\end{equation}
The subscript $C$ denotes that a central differencing scheme is used.

Eq.~\eqref{e:norm} can be broken down into 3 separate objectives, i.e.\ Eq.~\eqref{e:W}, \eqref{e:V}, and \eqref{e:G}. Denoting the matricisations of $\mathcal{J}$ along its rows, columns and tubes by $\boldsymbol{J}_{(1)}, \boldsymbol{J}_{(2)}, \boldsymbol{J}_{(3)}$, respectively, and using `$\kr$' to denote the Khatri-Rao product, an update of $\boldsymbol{W}$ is found from
\begin{equation}
\label{e:W}
\underset{\boldsymbol{W}}{\operatorname{arg~min}}~\Vert \boldsymbol{J}_{(1)} - \boldsymbol{W}((\mathcal{F}_C(\boldsymbol{V}) \circ \boldsymbol{G}) \kr \boldsymbol{V})^{\text{T}} \Vert_F^2
\end{equation}
Given that $\boldsymbol{W}$ appears linearly in the objective, a closed-form update formula may be obtained.

What is imperative is to retrieve meaningful estimates of the functions $g_i$ along the columns of $\boldsymbol{G}$. Originally one relied on the uniqueness properties of the CPD to meet this requirement \citep{dreesen2014}. As was mentioned earlier, it has been shown that this uniqueness is often not guaranteed, leading to very noisy estimate. 

The F-CPD method, on the other hand, relies on finite difference filters to steer the decomposition towards meaningful factors. What is required in practice is to promote smoothness on the estimates stored in $\boldsymbol{G}$. Given that both $\boldsymbol{V}$ (via $z_i = \boldsymbol{v}_i^{\text{T}} \boldsymbol{p}$) and $\boldsymbol{G}$ contribute to the smoothness of the estimates of $g_i(z_i$), a smoothness objective should be reflected in both their cost functions. Regularisation is used to penalise noisy estimates.  An update of $\boldsymbol{V}$ is found from
\begin{equation}
\label{e:V}
\begin{split}
\underset{\boldsymbol{V}}{\operatorname{arg~min}}&~\Vert \boldsymbol{J}_{(2)} - \boldsymbol{V}((\mathcal{F}_C(\boldsymbol{V}) \circ \boldsymbol{G}) \kr \boldsymbol{W})^{\text{T}} \Vert_F^2 \\
&+ \lambda \Vert \left(\mathcal{F}_L(\boldsymbol{V}) \circ \boldsymbol{G} \right) - \left(\mathcal{F}_R(\boldsymbol{V}) \circ \boldsymbol{G} \right) \Vert_F^2
\end{split}
\end{equation}
where $\lambda$ is a hyperparameter which balances both objectives. The additional term penalises divergent results from a left ($\mathcal{F}_L$) and a right ($\mathcal{F}_R$) finite difference filtering operation, ultimately steering the optimisation towards smooth solutions. Given that $\boldsymbol{V}$ appears nonlinearly in Eq.~\eqref{e:V}, nonlinear optimisation is required when computing an update.

In analogy with Eq.~\eqref{e:V}, the update formula of $\boldsymbol{G}$ is also found from a joint objective function.
\begin{equation}
\label{e:G}
\begin{split}
\underset{\boldsymbol{G}}{\operatorname{arg~min}}&~\Vert \boldsymbol{J}_{(3)} -(\mathcal{F}_C(\boldsymbol{V}) \circ \boldsymbol{G})(\boldsymbol{V} \kr \boldsymbol{W})^{\text{T}} \Vert_F^2 \\
&+ \lambda \Vert \left(\mathcal{F}_L(\boldsymbol{V}) \circ \boldsymbol{G} \right) - \left(\mathcal{F}_R(\boldsymbol{V}) \circ \boldsymbol{G} \right) \Vert_F^2
\end{split}
\end{equation}
It can be shown that $\boldsymbol{G}$ appears linearly in Eq.~\eqref{e:G}.

The smoothness objective ensures that the functions may be parameterised using an appropriate basis expansion. What basis expansion to use depends on the application and can be freely chosen by the user. 

\revv{The computational cost of the algorithm is quadratic in both $N$ and $r$. More efficient implementations are the subject of future study. For now, the benchmark problem of Section \ref{s:benchmark} could be solved in a computing time in the order of minutes. The present algorithm can be summarised by the following pseudocode.}

\rev{
\begin{table}[h!]
%\caption{Relative root-mean-squared error of function approxiation}
\label{t:1}
\begin{center}
\begin{tabular}{l}
\hline
\revv{\textbf{F-CPD Algorithm}} \\
\hline
\revv{\textbf{Construct} Jacobian tensor $\mathcal{J}$} \\
\revv{$\quad$ Randomly select $N$ points from the input space $\left\{ \boldsymbol{p}[k]\right\}_{k=1}^N$} \\
\revv{$\quad$ Compute $\boldsymbol{J}_k$ on the operating points $\left\{ \boldsymbol{p}[k]\right\}_{k=1}^N$}  \\
\revv{$\quad$ Stack the Jacobian matrices $\mathcal{J}[:,:,k] \coloneqq \boldsymbol{J}_k$} \\
\revv{\textbf{Factor} $\mathcal{J}$ into $\left\{ \boldsymbol{W},\boldsymbol{V}, \boldsymbol{G}\right\}$}\\
\revv{$\quad$ Initialise $\boldsymbol{W},\boldsymbol{V}, \boldsymbol{G}$ for chosen value of $r$}\\
\revv{$\quad$ \textbf{Repeat} ALS routine}\\
\revv{$\quad$ update $\boldsymbol{W},\boldsymbol{V}, \boldsymbol{G}$ via \eqref{e:W} to \eqref{e:G}}\\
\revv{$\quad$ \textbf{Until} maximum number of iterations}\\
\revv{\textbf{Parameterise} $\boldsymbol{G}$}\\
\revv{$\quad$ obtain $g_i(z_i)$ from appropriate basis expansion of $\boldsymbol{G}$}\\
\hline
\end{tabular}
\end{center}
\end{table}}

The filtered CPD approach is a powerful tool since it no longer relies on the uniqueness properties of the CPD. As a result, the number of univariate functions, $r$, to be used in the decoupled function, has become a design choice. 

Moreover, also multiple-input single-output functions, may be decoupled using filtered CPD. MISO functions result in Jacobian matrices rather than tensors, preventing the possibility of exploiting the uniqueness properties of the CPD. F-CPD therefore enables the decoupling of single-output NARX models, and by extent of general MISO functions. %At the core of the algorithm lies an alternating least-sqaures routine which iteratively updates $\boldsymbol{W}, \boldsymbol{V}$, and $\boldsymbol{G}$. The method relies on finite difference operations (or filters) to steer the decomposition towards meaningful factors. What is required in practice is to promote smoothness on the non-parametric estimates stored in $\boldsymbol{G}$. Regularisation is used to penalise noisy estimates, ensuring that the functions may be parameterised using an appropriate basis expansion. What basis expansion to use depends on the application and can be freely chosen by the user. The filtered CPD approach is a powerful tool since it no longer relies on the uniqueness properties of the CPD. As a result, the number of univariate functions, $r$, to be used in the decoupled function, has become a design choice. Moreover, also multiple-input single-output functions (MISO), may be decoupled using filtered CPD. MISO functions were originally excluded since they result in Jacobian matrices rather than tensors, losing the required uniqueness properties of tensor decompositions. Overcoming this obstacle enables the decoupling of single-output NARX models.

%\begin{figure}
%\begin{center}
%\begin{tikzpicture}
%\draw [thick] (0,0) rectangle (5,5);
%\node at (0,0) {(0,0)};
%\end{tikzpicture}
%\caption{}
%\label{f:algorithm}
%\end{center}
%\end{figure}

\section{Benchmark problem}
\label{s:benchmark}
The decoupling of single-output P-NARX models will be demonstrated on a nonlinear benchmark data set of the forced duffing oscillator. The data is obtained from an electrical implementation of a mechanically resonating system involving a moving mass $m$, a viscous damping $c$ and a nonlinear spring $k(y(t))$. The analogue electrical circuitry generates data close to but not exactly equal to the idealised representation given by the nonlinear ordinary differential equation (ODE)
\begin{equation}
\label{e:diff_SB}
%m \frac{d^2\textbf{y}(t)}{dt} + c \frac{d\textbf{y}(t)}{dt} + k(\textbf{y}(t))\textbf{y}(t) = \textbf{u}(t),
m \ddot{y}(t) + c \dot{y}(t) + k(y(t))y(t) = u(t),
\end{equation}
where the presumed displacement, $y(t)$, is considered the output and the presumed force, $u(t)$, is considered the input. Overdots denote the derivative with respect to time. The static position-dependent stiffness is given by
\begin{equation}
\label{e:SB_kNL}
k(y(t))=\alpha+\beta y^2(t),
\end{equation}
which can be interpreted as a cubic hardening spring.

\textbf{The training data} consists of 9 realisations of a random-phase odd multisine.
% signal given by
%%
%\begin{equation}
%\label{e:multisine}
%u(t) = A \sum_{\begin{matrix} l=1\\l~odd \end{matrix}}^{L}\cos \left( 2\pi f_0 \ell t+ \phi_l \right),
%\end{equation}
%%
The period of the multisine is $\rfrac{1}{f_0}$ with $f_0= \rfrac{f_s}{8192}$ Hz and $f_s\approx 610$ Hz. The number of excited harmonics is $L=1342$ resulting in an $f_{\text{max}}\approx 200$ Hz. Each multisine realisation is given a unique set of phases that are independent and uniformly distributed in $[0,2\pi[$. The signal to noise ratio (SNR) at the output is estimated at approximately 40 dB. \revv{This is measurement noise (or sensor noise). It is high levels of measurement noise can deteriorate the performance of NARX models \citep{schoukens2021}. It is important to stress that this inherent sensitivity is not removed by replacing the internal function by a decoupled one.}

As \textbf{validation data} a filtered Gaussian noise sequence of the same band width and with a linearly increasing amplitude is used. 

The data are part of three benchmark data sets for nonlinear system identification described in \cite{wigren2013}.
The approach will consist of 3 steps:
\begin{enumerate}
\item Identify a P-NARX model on the basis of the training data.
\item Decouple the NARX model using the F-CPD technique.
\item Use a final optimisation to minimise the simulation error of the decoupled model.
\end{enumerate}

\subsection{Reference P-NARX model}
\label{s:PNARX}
Using the System Identification toolbox in MATLAB a P-NARX model is estimated with the following properties: \revv{$n_u = 1$}, $n_y = 3$, and $d = 3$. All cross-term monomials where included leading to a model with 55 parameters. We will denote the P-NARX model by $f(\boldsymbol{x}): \mathbb{R}^{n_u + n_y +1} \rightarrow \mathbb{R}$, with $\boldsymbol{x}$ as defined in Eq.~\eqref{e:x}, \rev{emphasising that the model is obtained by minimising the equation error (focus on prediction)}. As performance metric a relative root-mean-squared simulation error is used
 \begin{equation}
 \label{e:rel_er}
 e_{\text{rms}} = \frac{\sqrt{\frac{1}{N}\sum_{k=1}^N\left(y[k]-y_s[k]\right)^2}}{\sqrt{\frac{1}{N}\sum_{k=1}^N y[k]^2}} \times 100
 \end{equation}
 where $y_s$ denotes the simulated output and $N$ is the record length. The estimation process returns an accurate P-NARX model yielding a $e_{\text{rms}}$ of 1.01\% when simulating the validation data.

 \subsection{Decoupled P-NARX model}
 \label{s:FCPD}
 Given that using the F-CPD method, $r$, has become a design choice, we are able to scan over $r$ and study the performance of the obtained decoupled models. Additionally, a scan over the hyperparameter $\lambda$ is required. Besides the simulation error (Eq.~\eqref{e:rel_er}) we will also introduce a function approximation error,
 \begin{equation}
 \label{e:rel_f}
e_f = \frac{\sqrt{\frac{1}{N}\sum_{k=1}^N\left(f(  \color{black} \boldsymbol{x}_o[k]   \color{black})-f_{d}(  \color{black} \boldsymbol{x}_o[k]   \color{black})\right)^2}}{\sqrt{\frac{1}{N}\sum_{k=1}^N f(  \color{black}\boldsymbol{x}_o[k]   \color{black})^2}} \times 100
 \end{equation}
 in which $f_d$ represents the decoupled polynomial function,
 \begin{equation}
 \label{e:fd}
 f_d = \boldsymbol{w}^{\text{T}} \boldsymbol{g} \left(\boldsymbol{V}^{\text{T}}  \color{black} \boldsymbol{x}_o   \color{black} \right),
  \end{equation}
  \rev{and $\boldsymbol{x}_o$ are the operating points.}
  In this case $N$ refers to the number of operating points for which the decomposition is computed.

In order to ensure that the operating points cover the region of interest, they are selected on the basis of the training data. \rev{Simulating the training data using the P-NARX estimate $f$ returns the collection $\boldsymbol{X}_T = \{\boldsymbol{x}_T[k]\}_{k=0}^{N_T}$, with $\boldsymbol{x}_T$ containing the simulated output samples}
\begin{multline}
\label{e:xT}
\color{black} \boldsymbol{x}_T = \{u(t), u(t-1), \dots, u(t-n_u), \\ 
 \color{black} y_s(t-1), \dots, y_s(t-n_y)\}, \quad \quad
\end{multline}
and $N_T$ denoting the training record length. The operating points are then drawn from the joint normal distribution, inferred from $\boldsymbol{X}_T$. The number of operating points to use can be considered a hyperparameter. In this work $N = 200$ points is used. In correspondence to the reference model, $f$, the functions $g_i$ of the decoupled models will be parametrised using third order polynomials, $d=3$.

\begin{figure}
\begin{center}
\includegraphics[scale=0.3]{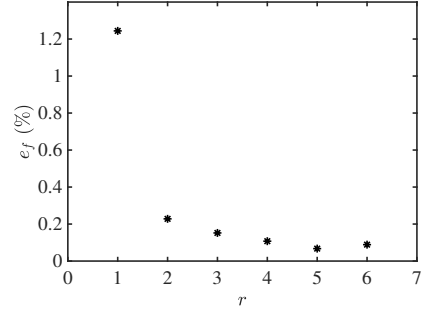}
\caption{Relative function approximation error computed on the operating points.}
\label{f:ef}
\end{center}
\end{figure}

We will study the decoupled models following the grid defined by $r=1, \dots, 6$ and $\lambda = 10^l$, with $l=-1,\dots, 5$. For every value of $r$ the decoupled model yielding the lowest function approximation error, $e_f$, is selected. This will correspond to the value of $\lambda$ which results in the appropriate balance between the tensor approximation objective and the smoothness objective (Eq.~\eqref{e:V} and \eqref{e:G}). The overview of the function approximation error is depicted in Fig.~\ref{f:ef}. Accurate function approximations, $f_d \approx f$, could be obtained yielding errors below 1\% for values of $r>1$.

In a final step, nonlinear optimisation is used to fine-tune the decoupled models on the basis of their \rev{training data simulation error}. A Levenberg-Marquardt algorithm \revv{\citep{levenberg1944}} is used to minimise the output error objective given by
\begin{equation}
\color{black} \underset{\boldsymbol{w},\boldsymbol{V},\bm{\theta}}{\operatorname{arg~min}} \sum_{k=1}^{N_T}\left( y[k]  - \boldsymbol{w}^{\text{T}} \boldsymbol{g}(\boldsymbol{V}^{\text{T}} \boldsymbol{x}_s[k],\bm{\theta})\right)^2,
\end{equation} 
\rev{with $y$ the true output, $\boldsymbol{x}_s$ containing simulated output samples of the decoupled model (similar to Eq.~\eqref{e:xT})}, and $\bm{\theta}$ storing the coefficients of the third order polynomials in $\boldsymbol{g}$.
 
\begin{figure}
\begin{center}
\includegraphics[scale=0.3]{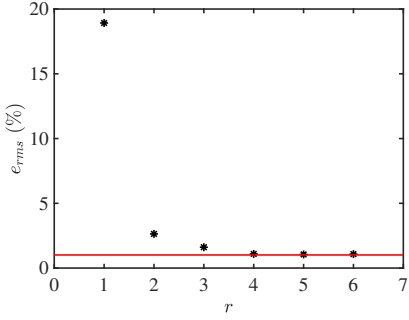}
\caption{Black markers: Relative simulation error of the decoupled P-NARX models, computed on the validation data. Red: error obtained from the reference P-NARX model.}
\label{f:eval}
\end{center}
\end{figure}

\begin{figure}
\begin{center}
\includegraphics[scale=0.3]{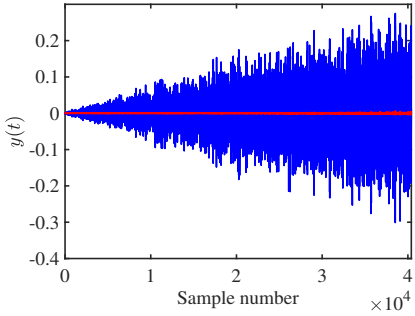}
\caption{Blue: validation output data corresponding to a filtered Gaussian noise sequence. The amplitude extends beyond the training data amplitude resulting in an extrapolation of the model. Red: simulation error of the $r=4$ decoupled model.}
\label{f:r4_val}
\end{center}
\end{figure}

Fig.~\ref{f:eval} illustrates the performance of the optimised model set when simulating the validation data. It is clear that decoupled models for which $r\ge4$ perform equally well as the reference model, i.e\ $\approx1$\% (indicated by the red line). Models containing less univariate mapping functions miss the required complexity to reproduce the data up to such precision. For some applications, however, the performance of the $r=2$ and $r=3$ model may still be acceptable. The simulation performance of the $r=4$ model is illustrated in Fig.~\ref{f:r4_val}.

\begin{figure}
\begin{center}
\includegraphics[width=0.5\textwidth]{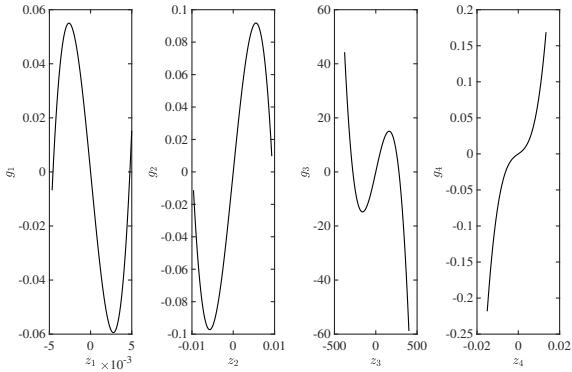}
\caption{Univariate functions of the $r=4$ decoupled P-NARX model. The functions are evaluated over the validation range.}
\label{f:r4_g}
\end{center}
\end{figure}

Noticing that $\boldsymbol{w}$ merely serves as a scaling on the functions $\boldsymbol{g}$, the coefficients can be incorporated in $\bm{\theta}$. Doing so results in a total parameter count of 36 for the $r=4$ model, \revv{compared to the 55 parameters required in the standard basis expansion of the reference model}. The univariate functions of the decoupled model are depicted in Fig.~\ref{f:r4_g}. \revv{From visual inspection of the functions one may conclude that the system behaviour is dominantly cubic. Additionally the SISO functions can easily be monitored to flag extrapolation \citep{karami2019}. It should be noted that in this case, even after decoupling, physical interpretability remains hard. An important aspect is believed to be the choice of regressors, i.e.\ $n_u,n_y$. The use of shifted input samples contributes to retrieving a non-physical model, in this case of the forced Duffing oscillator. One such mechanism is the introduction of sampling zeros \citep{goodwin2013}.} 

\section{Comparing polynomial constraints to F-CPD}
In this section the decoupling technique of \cite{karami2021}, which will be referred to as the \emph{the structured Hessian method}, is compared to the F-CPD method. \rev{Following the author's notation, the decoupled model may be written as
\begin{subequations}
\label{e:fdk}
\begin{align}
&f_d = c_0 + \sum_{i=1}^r g_i(\boldsymbol{v}_i^{\text{T}} \boldsymbol{x})\\
&g_i(z_i) = \sum_{j=1}^d c_{j,i} z_i^j,
\end{align}
\end{subequations}
which corresponds to Eq.~\eqref{e:fd} where $\boldsymbol{w}$ is incorporated in the vector of coefficients,
\begin{equation}
\boldsymbol{c} = [ c_0~c_{1,1}~\dots~c_{d,1}~c_{1,2}~\dots~c_{d,r}]^{\text{T}}.
\end{equation}
Computing the Hessian of the output of the decoupled function with respect to the input variables generates a matrix for each operating point $\boldsymbol{x}_o$
\begin{equation}
 \boldsymbol{H}'_y(\boldsymbol{x}_o) = \sum_{i=1}^r g_i''(\boldsymbol{v}_i^{\text{T}} \boldsymbol{x}_o) \boldsymbol{v}_i \boldsymbol{v}_i^{\text{T}}
 \end{equation}
where $g_i''$ denotes the second derivative of $g_i$ with respect to $z_i$. The collection of Hessian matrices may again be stacked into a three-way array whose entries are given by
\begin{equation}
\mathcal{H}'_{[j,k,l]} = \sum_{i=1}^r v_{ij}~v_{ik}~g_{il}''
\end{equation}
$\text{for}~j=1,\dots,n~\text{and}~k=1,\dots,n~\text{and}~l=1,\dots,N$.

The idea is to construct the Hessian tensor $\mathcal{H}$ out of evaluations of the known function and use the CPD to factor it into $\mathcal{H}' = \llbracket \boldsymbol{V}, \boldsymbol{V}, \boldsymbol{G}'' \rrbracket$, such that $\mathcal{H} \approx \mathcal{H}'$. In this case $\boldsymbol{G}'' = [\boldsymbol{g}_1'' \dots \boldsymbol{g}_r'']$ is a matrix storing evaluations of the second derivate of $\boldsymbol{g}$.

From the previous it is clear that relying on the uniqueness properties of the CPD alone does not guarantee to obtain accurate estimates of $g_i''$. It was therefore proposed not to solve for $\boldsymbol{G}''$ directly, but to formulate the matrix factor into a polynomial form and solve for the polynomial coefficients instead. The optimisation problem to be solved can be formulated as
\begin{equation}
\label{e:GP}
\underset{\boldsymbol{V},\boldsymbol{c}}{\operatorname{arg~min}} \lVert \mathcal{H}  - \llbracket \boldsymbol{V},\boldsymbol{V},\boldsymbol{G}''_P(\boldsymbol{c},\boldsymbol{V}) \rrbracket \rVert_F^2,
\end{equation} 
where the columns of $\boldsymbol{G}''_P$ are expressed as polynomials using the coefficients $\boldsymbol{c}$ and the Vandermonde matrices based on all the $z_i$ (hence the dependence on $\boldsymbol{V}$ given $z_i = \boldsymbol{v}_i^{\text{T}}\boldsymbol{x}$). The optimisation problem is solved using a quasi-Newton algorithm.

The obtained estimate of $\boldsymbol{V}$ is then used to initialise a final optimisation based on the simulated training output of the decoupled model
\begin{equation}
\underset{\boldsymbol{V},\boldsymbol{c}}{\operatorname{arg~min}}  \sum_{k=1}^{N_T}\left( y[k]  - y_s[k](\boldsymbol{V},\boldsymbol{c})\right)^2.
\end{equation} 
 }

Starting from the reference model of Section \ref{s:PNARX}, the structured Hessian method is used to obtain a decoupled P-NARX model. The exact same data and operating points are used for fair comparison. 

\begin{figure}
\begin{center}
\includegraphics[scale=0.3]{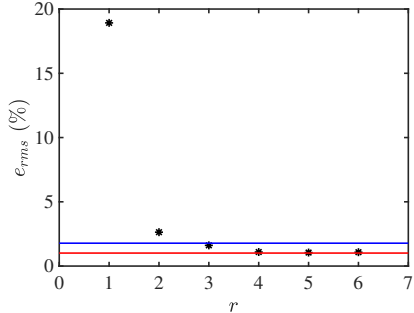}
\caption{Black markers: Relative simulation error of the decoupled P-NARX models, computed on the validation data. Red: error obtained from the reference P-NARX model. Blue: error of the $r=4$ model obtained from the structured Hessian method.}
\label{f:hessian}
\end{center}
\end{figure}

A decoupled P-NARX model with $r=4$ functions is obtained. Also here polynomials of the third degree are used. The obtained model is of an identical architecture as the $r=4$ model obtained in Section \ref{s:FCPD}, i.e.\ when considering the $\boldsymbol{w}$ vector to be incorporated in $\boldsymbol{g}$. In Fig.~\ref{f:hessian}, the performance of the decoupled model is compared to the models obtained in Section \ref{s:FCPD}. The decoupled model has a slightly higher error of \rev{$e_{\text{rms}} = 1.78$\%}.

A number of advantages and disadvantages of both methods are listed:
\begin{itemize}
\item The structured Hessian method requires the expensive computation of the Hessian.
\item The F-CPD method requires a search over the hyperparameter $\lambda$.
\item The structured Hessian method enforces polynomial constraints on one of the factors. This may be inappropriate when facing non-polynomial nonlinearities.
\item \rev{The structured Hessian method retrieves an estimate of the factor $\boldsymbol{V}$ from the constrained CPD of the Hessian. The Hessian, however, no longer contains information on the linear part of the NARX model.}
\end{itemize}
\section{Conclusion}
In this work a filtered CPD is used to decouple single-output polynomial NARX models. Polynomial NARX models typically require a large number of parameters in their description. For decoupled structures, the number of parameters grows linearly with the degree, resulting in a substantial model reduction. It was shown that the filtered CPD method no longer relies on the uniqueness properties of the CPD. As a result, the number of univariate functions in the decoupled structure has become a design choice. The method is illustrated on the benchmark problem of the forced Duffing oscillator and compared to the results obtained from the structured Hessian method. 

\bibliography{entirebib}             % bib file to produce the bibliography
                                                     % with bibtex (preferred)
                                                   
%\begin{thebibliography}{xx}  % you can also add the bibliography by hand

%\bibitem[Able(1956)]{Abl:56}
%B.C. Able.
%\newblock Nucleic acid content of microscope.
%\newblock \emph{Nature}, 135:\penalty0 7--9, 1956.

%\bibitem[Able et~al.(1954)Able, Tagg, and Rush]{AbTaRu:54}
%B.C. Able, R.A. Tagg, and M.~Rush.
%\newblock Enzyme-catalyzed cellular transanimations.
%\newblock In A.F. Round, editor, \emph{Advances in Enzymology}, volume~2, pages
%  125--247. Academic Press, New York, 3rd edition, 1954.

%\bibitem[Keohane(1958)]{Keo:58}
%\newblock \emph{Power and Interdependence: World Politics in Transitions}.
%\newblock Little, Brown \& Co., Boston, 1958.

%\bibitem[Powers(1985)]{Pow:85}
%\newblock Is there a way out?
%\newblock \emph{Harpers}, pages 35--47, June 1985.

%\bibitem[Soukhanov(1992)]{Heritage:92}
%\newblock \emph{{The American Heritage. Dictionary of the American Language}}.
%\newblock Houghton Mifflin Company, 1992.

%\end{thebibliography}

\end{document}